\documentclass[aps,prb,reprint,superscriptaddress,intlimits]{revtex4-2}

\usepackage{bm,latexsym,mathrsfs,enumerate,amsmath,amssymb}

\usepackage[breaklinks=true,unicode=true,urlcolor = blue,colorlinks = true,citecolor = blue,linkcolor = blue]{hyperref}
\usepackage{graphicx}

\renewcommand{\vec}[1]{\bm{#1}}

\newcommand{\md}{\mathrm{d}}

\begin{document}
	
\title{Instability of magnetic skyrmion strings induced by longitudinal spin currents}

\author{Shun Okumura}
\affiliation{The Institute for Solid State Physics, The University of Tokyo, Kashiwa, Chiba, 277-8581, Japan}

\author{Volodymyr P. Kravchuk}
\affiliation{Institut f\"ur Theoretische Festk\"orperphysik, Karlsruher Institut f\"ur Technologie, D-76131 Germany}
\affiliation{Bogolyubov Institute for Theoretical Physics of National Academy of Sciences of Ukraine, 03143 Kyiv, Ukraine}

\author{Markus Garst}
\affiliation{Institut f\"ur Theoretische Festk\"orperphysik, Karlsruher Institut f\"ur Technologie, D-76131 Germany}
\affiliation{Institute for Quantum Materials and Technology, Karlsruhe Institute of Technology, D-76131 Karlsruhe, Germany}

\begin{abstract}
It is well established that spin-transfer torques exerted by in-plane spin currents give rise to a motion of magnetic skyrmions resulting in a skyrmion Hall effect. In films of finite thickness or in three-dimensional bulk samples the skyrmions extend in the third direction forming a string. We demonstrate that a spin current flowing longitudinally along the skyrmion string instead induces a Goldstone spin wave instability. Our analytical results are confirmed by micromagnetic simulations of both a single string as well as string lattices suggesting that the instability eventually breaks the strings. A longitudinal current is thus able to melt the skyrmion string lattice via a dynamical phase transition. For films of finite thickness or in the presence of disorder a threshold current will be required, and we estimate the latter assuming weak collective pinning.   
\end{abstract}

\maketitle

One of the fascinating aspects of two-dimensional topological spin textures, so-called magnetic skyrmions, 
is their interplay with spin currents \cite{Nagaosa2013}. When electrons adiabatically traverse a skyrmion and locally adjust their spin degree of freedom, they will be influenced by an emergent orbital magnetic field which is proportional to the topological winding number of the texture \cite{Volovik1987}. 
The resulting emergent Lorentz force is at the origin of the topological Hall effect \cite{Bruno2004,Neubauer2009,Lee2009}. Vice versa, a spin current also exerts a force on skyrmions that gives rise to a skyrmion Hall effect. Early experiments demonstrated that a lattice of magnetic skyrmions in bulk chiral magnets like MnSi and FeGe can be manipulated by currents on the order of an ultralow threshold of $10^6$ A/m$^2$ \cite{Jonietz2010,Schulz2012,Yu2012}. This observation triggered many activities with the aim to exploit magnetic skyrmions for spintronic applications \cite{Fert2013,Iwasaki2013,Sampaio2013}. In thin magnetic films, skyrmions are now routinely manipulated and moved by spin currents that flow within the plane of the film   \cite{Jiang2015,Jiang2017,Woo2016,Litzius2017,Zazvorka2019,Litzius2020,Fert2017,Back2020}. 

\begin{figure}
\includegraphics[width=0.9\columnwidth]{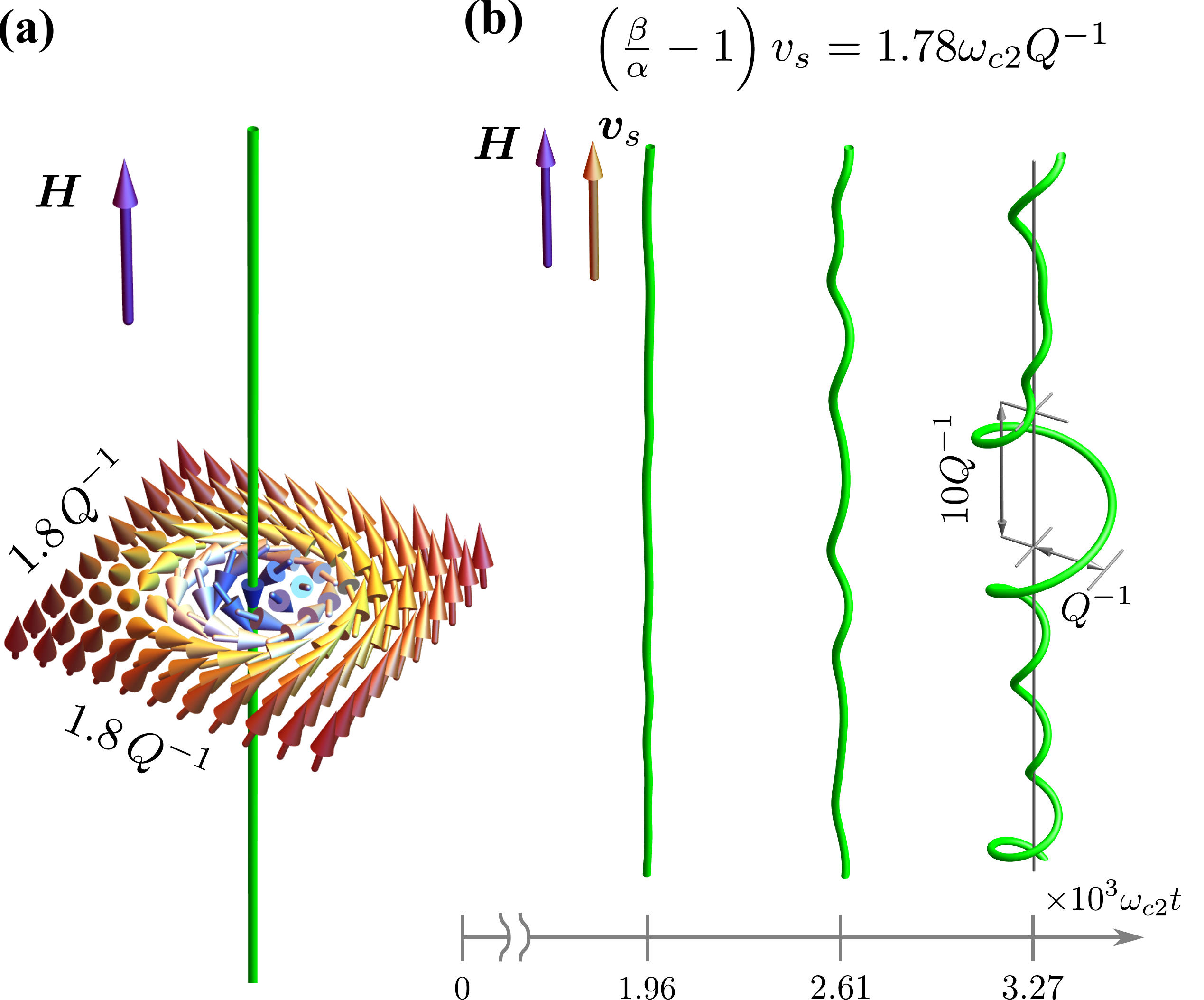}
\caption{(a) Illustration of a skyrmion string that is aligned with the applied magnetic field $H$; each cross section perpendicular to $H$ contains a skyrmion texture, and the green line represents the first-moment of topological charge. (b) Micromagnetic simulation of a string destabilized by a longitudinal spin current $v_s$; 
after initialization with small random fluctuations the amplitudes of translational Goldstone modes 
grow exponentially in time, for details see text and Ref.~\cite{SI}.
}
\label{fig1}
\end{figure}

In bulk materials, the two-dimensional skyrmion texture extends in the third direction forming a skyrmion string, see Fig.~\ref{fig1}(a). This string is aligned with the applied magnetic field and possesses a tension. Recent advances in  real-space magnetic imaging techniques succeeded to visualize skyrmion strings close to surfaces \cite{Park2014,Yu2020,Birch2020,Birch2022,Wolf2022} and via magnetic X-ray tomography even within the bulk \cite{Seki2022}. 
Similar to Kelvin waves on vortex filaments, spin waves can propagate along skyrmion strings \cite{Lin2019}, and this propagation is non-reciprocal as confirmed experimentally \cite{Seki2020}. Moreover, the non-linear elasticity of skyrmion strings was found to stabilize solitary waves \cite{Kravchuk2020} whose analogues for vortex filaments were discussed by Hasimoto already in the 1970ies \cite{Hasimoto1972}. 

The influence of a transversal spin current on skyrmion strings, i.e.,~a current that flows perpendicular to the string, has been also widely investigated. In analogy with the two-dimensional case, a transversal current density homogeneously applied along the whole string will set the string into motion at least in the absence of pinning by disorder. There exist theoretical \cite{Lin16} and experimental \cite{Kagawa17} evidences that pulses of such currents are able to break the strings via the creation of hedgehog singularities, i.e., emergent magnetic monopoles, that lead to a topological unwinding of strings. It has been argued that the complex dynamical response of skyrmion strings results in a non-reciprocal non-linear Hall effect \cite{Yokouchi18}. The influence of defects on the motion of skyrmion strings was studied in Ref.~\cite{Zang2011,Koshibae2019,Koshibae20,Reichhardt2022}, and their elasticity was shown to be important for an understanding of the depinning transition \cite{Bezvershenko22}.

In the present work, we investigate the influence of spin currents that flow parallel, i.e., longitudinal to the skyrmion string. We show that, remarkably, such a current component immediately destabilizes the string in a clean system. This instability is caused by the longitudinal current 
leading to the emission of translational Goldstone modes with finite wavevectors along the string, $k_z \neq 0$, in contrast to the transversal current that couples only to the Goldstone mode with $k_z = 0$. As a result, helical deformations develop, see Fig.~\ref{fig1}(b), whose amplitudes grow with time and eventually break the string. Employing an analytical stability analysis complemented by micromagnetic simulations, we demonstrate that a single string as well as a skyrmion string lattice are destabilized by this mechanism. 

Our starting point is the Landau-Lifshitz-Gilbert equation describing the magnetization dynamics supplemented by spin-transfer torques \cite{Zhang04,Tserkovnyak08}
\begin{equation}
\label{eq:LLG}
\left(\partial_t + \vec{v}_s 
\vec{\nabla}\right)\vec{n} =
- \gamma\, \vec{n}\times \vec B_{\rm eff} 
+\alpha\,  \vec{n}\times \left(\partial_t + \frac{\beta}{\alpha} \vec{v}_s 
\vec{\nabla}\right) \vec{n} .
\end{equation}
The continuous unit vector field $\vec{n}=\vec{n}(\vec{r},t)$ specifies the local orientation of the magnetization, $\gamma$ is gyromagnetic ratio, $\alpha > 0$ is the Gilbert damping, and $\beta$ is the dissipative spin-torque parameter. The spin-transfer torques involve the effective spin velocity $\vec{v}_s$ that is parallel to the applied spin current density. The effective magnetic field, $\vec B_{\rm eff} = - \frac{1}{M_s} \frac{\delta V}{\delta \vec n}$ with the saturation magnetization $M_s$, is determined by the potential $V = \int d\vec{r} \mathcal{V}$. The potential density for a cubic chiral magnet in the limit of small spin-orbit coupling is given by 
\begin{align}
\label{eq:functional}
\mathcal{V} = A (\partial_i \vec n)^2 + D \vec n (\vec \nabla \times \vec n) - \mu_0 M_s H n_z,
\end{align}
with the exchange stiffness $A$, the Dzyaloshinskii-Moriya interaction $D$, the magnetic constant $\mu_0$, and the magnetic field $H$ applied along the $z$-axis. In the following, we choose a right-handed magnetic system $D>0$. 
We neglect dipolar interactions but comment on their influence in the supplementary information \cite{SI}. 
It is convenient to introduce the scales $\omega_{c2} = \gamma D^2/(2AM_s)$ and $Q = D/(2A)$ for frequency and wavevector, respectively, as well as the dimensionless magnetic field $h = \gamma \mu_0 H/\omega_{c2}$. 

For $h \geq 1$ the ground state of Eq.~\eqref{eq:functional} is field-polarized, and in the absence of  spin-transfer torques, $\vec{v}_s = 0$, the static skyrmion string $\vec{n}_{\rm SkS}(\vec r)$ of Fig.~\ref{fig1}(a) is a topologically stable excitation  \cite{Bogdanov94}. For a finite in-plane spin velocity, $\vec{v}_s \perp \hat{\vec{z}}$, the equation of motion is solved by a drifting skyrmion string $\vec{n}_{\rm SkS}(\vec r - \vec{v}_d t)$ with an in-plane drift velocity $\vec{v}_d \nparallel \vec{v}_s$ realizing a skyrmion Hall effect \cite{Nagaosa2013}. 

\begin{figure}
\includegraphics[width=0.9\columnwidth]{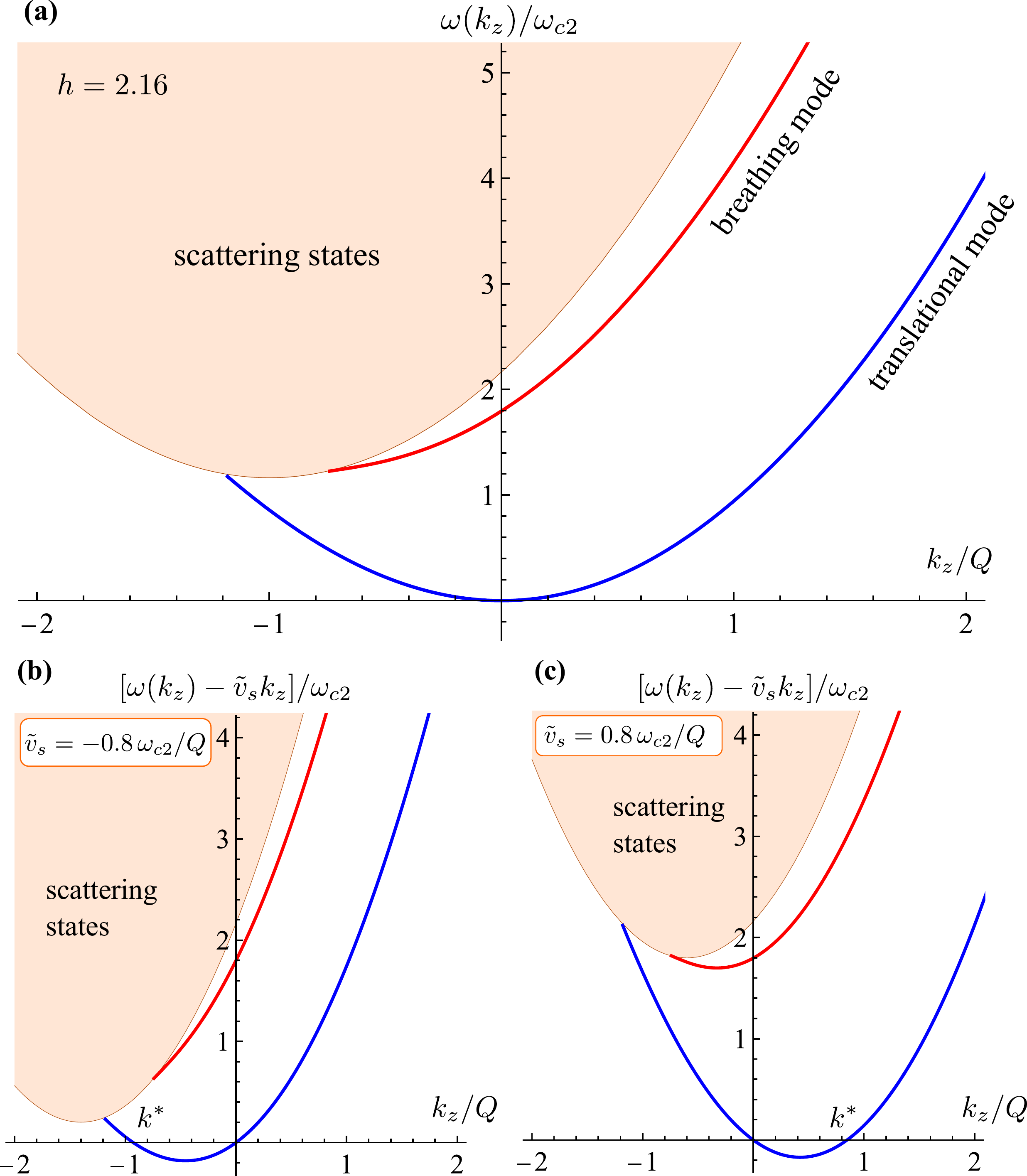}
\caption{ (a) Spin wave spectrum of a single skyrmion string at $h = 2.16$, see Refs.~\cite{Lin2019,Kravchuk2020} for details, consisting of the translational Goldstone mode, the breathing mode and extended scattering states. (b,c) In the presence of a small longitudinal current $v_s \neq 0$, the stability criterion \eqref{eq:criterion} is violated for the translational Goldstone mode with wavevectors from zero to $k^*$ for which $\omega(k_z) - \tilde{v}_s k_z < 0$ where $\tilde{v}_s \equiv (\beta/\alpha - 1)v_s $.}
\label{fig2}
\end{figure}

Here, we are interested, however, in a longitudinal spin current $\vec{v}_s = v_s \hat{\vec{z}}$; as the spin-transfer torques act on the magnetization only via the operator $v_s \partial_z$ and $\partial_z \vec{n}_{\rm SkS}(\vec r) = 0$, the static skyrmion string $\vec{n}_{\rm SkS}(\vec r)$ still solves Eq.~\eqref{eq:LLG}. In order to determine the dynamical stability of the static string solution for $v_s \neq 0$ we perform a stability analysis by examining small fluctuations. Treating the Gilbert damping $\alpha$ on the right-hand side of Eq.~\eqref{eq:LLG} perturbatively, the complex fluctuation spectrum is obtained with the help of linear spin-wave theory, see Ref.~\cite{SI} for details,
\begin{align}
\label{eq:spectrum}
&\omega_{v_s}(k_z) =  \\\nonumber
&\qquad\frac{\beta}{\alpha} v_s k_z + \left(\omega(k_z) - \left(\frac{\beta}{\alpha} - 1\right)  v_s k_z\right)\left(1 - i \alpha P(k_z)\right).
\end{align}
Here, $k_z$ is the wavevector of the spin wave fluctuation, and the dimensionless function $P(k_z) \geq 1$ parametrizes its ellipticity \cite{Kambersky1975,Rozsa2018}. For $v_s = 0$ and $\alpha = 0$, the spectrum reduces to the spin wave dispersion $\omega(k_z)$ for the skyrmion string in equilibrium that was determined before by Lin {\it et al.} \cite{Lin2019}, see Fig.~\ref{fig2}(a).

The skyrmion string is stable as long as Im $\omega_{v_s}(k_z) < 0$ so that the factor $e^{- i \omega_{v_s}(k_z) t}$ accompanying the fluctuation amplitude decays exponentially in time. From Eq.~\eqref{eq:spectrum}
follows the stability criterion
\begin{equation}
\label{eq:criterion}
\omega(k_z) -   \left(\frac{\beta}{\alpha} - 1\right) v_s k_z > 0,
\end{equation}
that generalizes and agrees with previous work \cite{Bazaliy1998,Fernandez-rossier2004,Tserkovnyak06,Kravchuk14c}. 
As the string possesses a translational Goldstone mode with a quadratic spectrum for small $k_z$, 
$\omega(k_z) \approx \mathcal{D} k_z^2$, with the stiffness $\mathcal{D}$, there exist for any value of $v_s$  a range of wavevectors $k_z$ for which the corresponding stability condition is not fulfilled, see Fig.~\ref{fig2}(b) and (c). This implies that the string in the presence of a small longitudinal current  is destabilized by the spontaneous emission of Goldstone spin waves. 

\begin{figure}
\includegraphics[width=0.9\columnwidth]{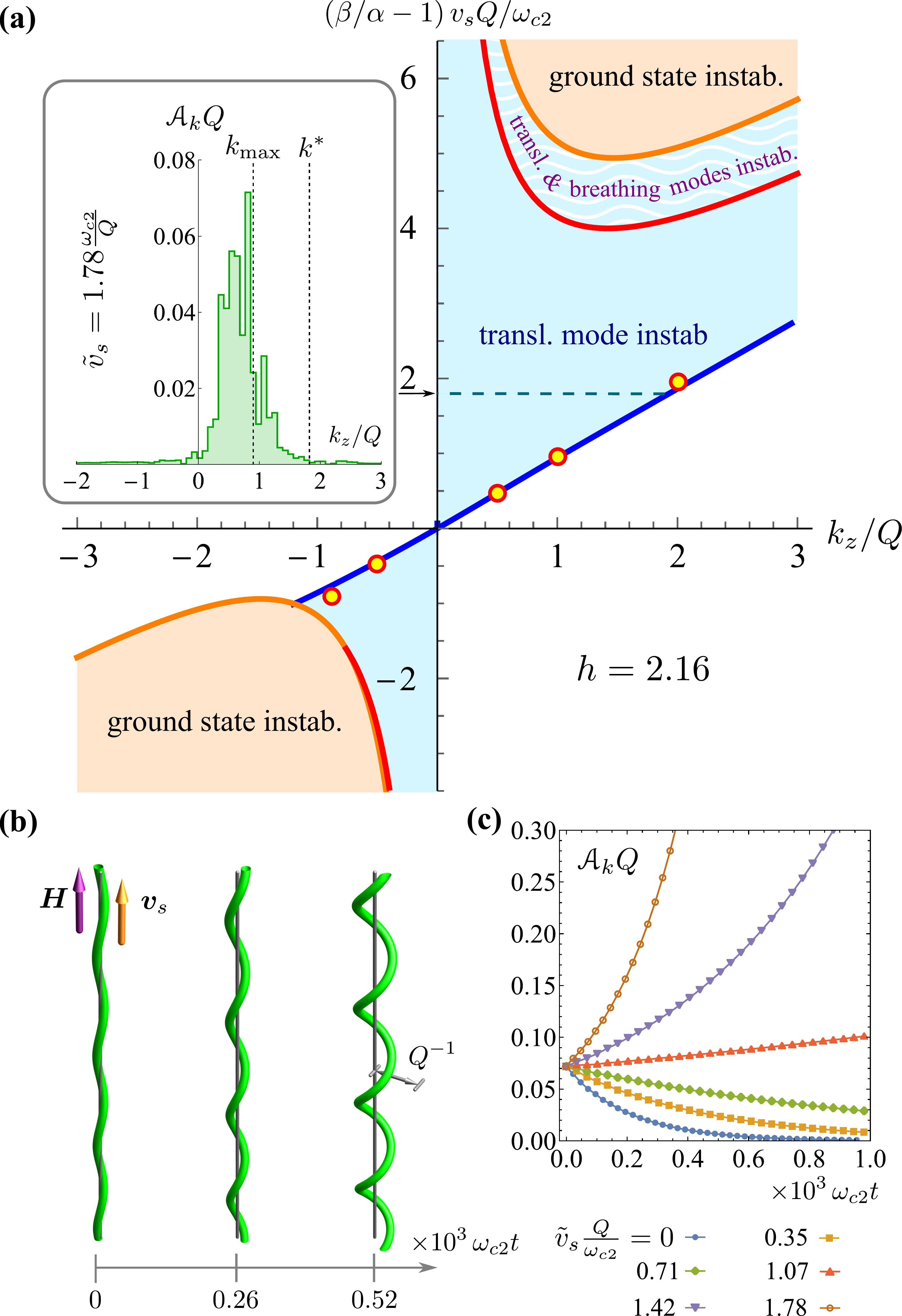}
\caption{(a) Instability diagram for the skyrmion string obtained with the help of Eq.~\eqref{eq:criterion} for a dimensionless magnetic field $h = 2.16$ and $\tilde{v}_s \equiv (\beta/\alpha - 1)v_s$. In the blue shaded region the translational Goldstone mode is unstable. The breathing mode of the string becomes unstable for larger $\tilde{v}_s$ at the red solid line. In the orange shaded region the field-polarized ground state is destabilized. Inset: Amplitude spectrum of the micromagnetic simulation displayed in Fig.~\ref{fig1}(b), see text. (b) Micromagnetic simulations of a string initialized with a Goldstone mode at wavevector $k_z = Q$. (c) Time evolution of the corresponding amplitude that allows to identify the critical value for $\tilde{v}_s$ indicated as yellow symbols in (a), see text.
}
\label{fig3}
\end{figure}

With this insight, we construct from the spectrum $\omega(k_z)$ of Fig.~\ref{fig2}(a) the instability diagram for the skyrmion string at the field $h=2.16$, see Fig.~\ref{fig3}(a); in Ref.~\cite{SI} the diagram is also shown for $h=1.08$. The blue shaded region bounded by the blue solid line indicates the range of wavevectors for which the translational Goldstone mode is unstable. In the limit of small $k_z$ the blue solid line is linear, $(\beta/\alpha- 1) v_s \approx \mathcal{D} k_z$, and determined by the stiffness $\mathcal{D}$ of the Goldstone mode. Increasing the values of $(\beta/\alpha- 1)v_s$  the breathing mode, in addition, becomes  unstable (red line) and, eventually, even the ground state (orange line). The latter occurs when the parabola bordering the scattering states in Fig.~\ref{fig2}(b) and (c) crosses zero, $((k_z+Q)/Q)^2 + h - 1 - k_z (\beta/\alpha- 1) v_s/\omega_{c2} =0$; it is the mode with $k_z = \pm \sqrt{h} Q$ that becomes unstable first at the velocity $(\beta/\alpha- 1) v_s|_{\rm min} = 2 (\pm \sqrt{h} + 1) \omega_{c2}/Q$ \cite{Lin2013}. Due to the non-reciprocity of the spectrum, Fig.~\ref{fig2}(a), the instability diagram lacks point symmetry with respect to the origin. 

 In order to validate Fig.~\ref{fig3}(a) we performed numerical micromagnetic simulations, see Ref.~\cite{SI} for details. The Goldstone amplitude spectrum of the numerically evaluated unstable skyrmion string shown in Fig.~\ref{fig1}(b) is analyzed in the inset of Fig.~\ref{fig3}(a). Indeed, amplitudes within the expected range of wavevectors from zero to $k^*$, with $\omega(k^*)/k^* = (\beta/\alpha- 1)v_s$, contribute. The imaginary part of $\omega_{v_s}(k_z)$ is maximal for a wavevector $k_{\rm max}$, for which the group velocity $\partial_{k_z}\omega(k_z)|_{k_{\rm max}} = (\beta/\alpha- 1)v_s$, and, as a consequence, this amplitude will develop the fastest. In agreement with this expectation, the amplitude spectrum has a peak close to $k_{\rm max}$. 
In order to verify the boundary of the instability region, i.e., the blue line in Fig.~\ref{fig3}(a), we deliberately excited the string in the micromagnetic simulations with a Goldstone mode of a given wavevector $k_z$ and monitored the time evolution of its amplitude, see Fig.~\ref{fig3}(b) and (c). Scanning various values of $(\beta/\alpha- 1)v_s$ we identified its critical value, for which the time evolution changes from an exponential decay to an exponential increase. The results for selected wavevectors are shown as yellow dots in panel (a) and are in good agreement with analytical predictions.
 
\begin{figure}
\includegraphics[width=0.9\columnwidth]{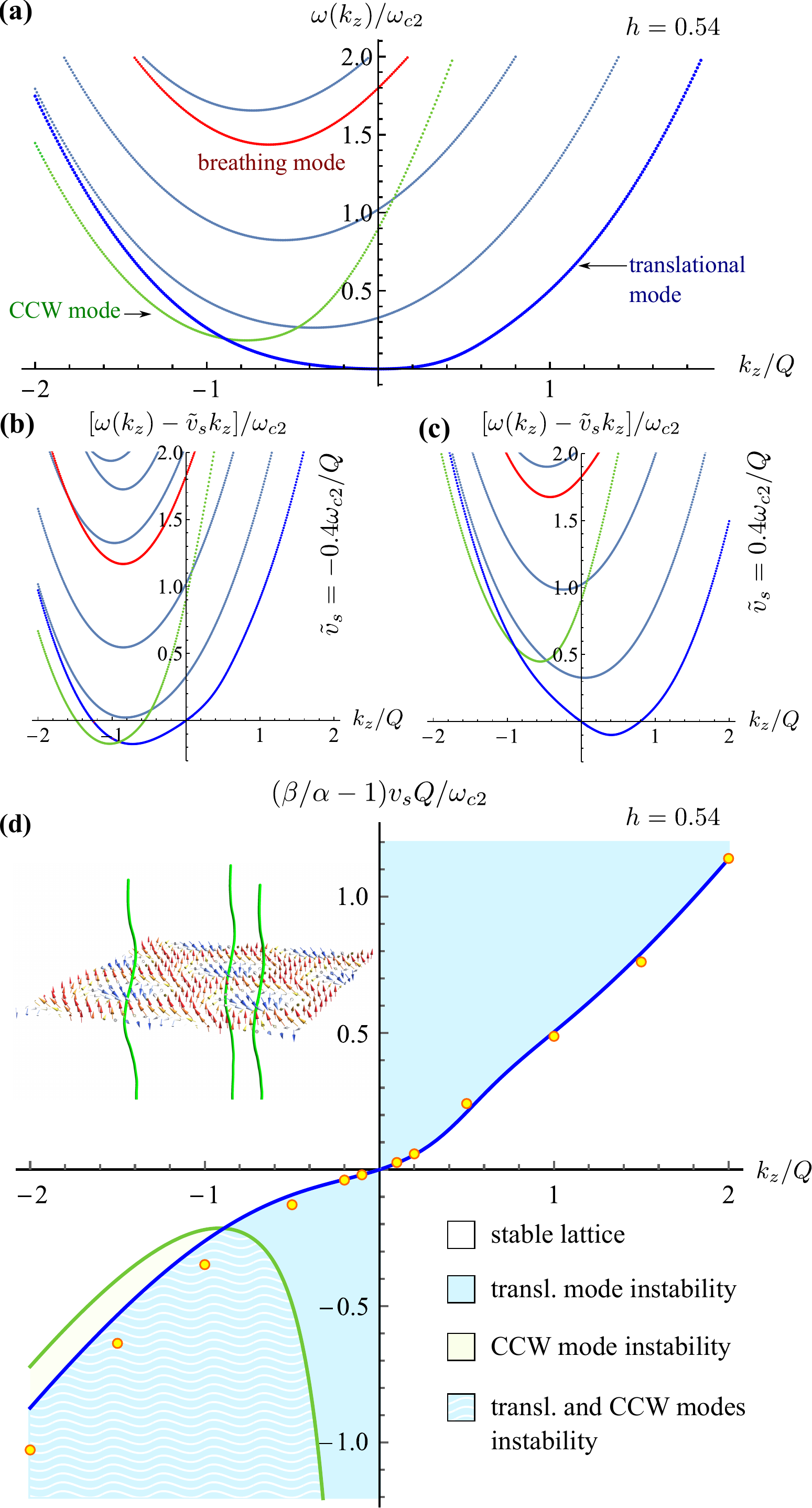}
\caption{(a) Magnon spectrum of the skyrmion string lattice at $h = 0.54$ for wavevectors $k_z$ along the strings \cite{Seki2020}. (b,c) Violation of the stability criterion \eqref{eq:criterion} in the presence of a finite $\tilde v_s \equiv (\beta/\alpha -1) v_s$. (d) Instability diagram for $h = 0.54$; in the blue shaded region the Goldstone mode is unstable. At the green solid line the CCW mode also destabilizes and, subsequently, various other modes (not shown). The yellow symbols are obtained via micromagnetic simulations, see inset and text.}
\label{fig4}
\end{figure}

Now, we turn to a discussion of skyrmion string lattices realized as a stable thermodynamic phase in cubic chiral magnets. Repeating the above arguments, we find that the stability criterion \eqref{eq:criterion} also applies to lattices where $\omega(k_z)$ should be  identified here with the spin wave spectrum at zero in-plane wavevectors, i.e., $\vec k = k_z \hat{\vec z}$. This spectrum was obtained in Ref.~\cite{Seki2020} and is shown in Fig.~\ref{fig4}(a). It comprises various modes \cite{Garst2017,Back2020} but the ones with lowest frequencies are the translational Goldstone mode and the counterclockwise (CCW) mode. Similar to the case of a single string, a small applied spin current immediately destabilizes the Goldstone mode of the skyrmion string lattice, see Fig.~\ref{fig4}(b) and (c). Increasing the spin current, the CCW mode, due to its pronounced non-reciprocity \cite{Seki2020}, is destabilized next for negative values of $(\beta/\alpha - 1) v_s$ and, subsequently, various other modes. This is summarized in the instability diagram of Fig.~\ref{fig4}(d). Note that the boundary of the Goldstone instability (blue solid line) is only linear $(\beta/\alpha- 1) v_s \propto k_z$ for very small $k_z \ll Q$ due to the pronounced non-reciprocity of the Goldstone dispersion in Fig.~\ref{fig4}(a).

Micromagnetic simulations were performed to confirm the diagram in Fig.~\ref{fig4}(d), for details see Ref.~\cite{SI}. Similar to Fig.~\ref{fig3}(b)and (c), the skyrmion string lattice was initialized with a Goldstone excitation with a specific wavevector, and the time evolution was monitored to determine the critical values of the velocity, that are shown as yellow symbols in Fig.~\ref{fig4}(d). There is very good agreement with the blue solid line obtained analytically except for larger negative values of $(\beta/\alpha - 1) v_s$. We attribute the slight deviations in this regime to the additional instability of the CCW mode that hinders a clear delineation of the Goldstone instability.

From the instability diagrams follows that even for very small spin currents the Goldstone spin wave instability is expected to develop for wavevectors $|k_z| \leq |(\beta/\alpha - 1) v_s|/\mathcal{D}$ where $\mathcal{D}$ represents the stiffness either of the single skyrmion string or the string lattice. The development will be however hampered if the Goldstone mode acquires an excitation gap. This is the case for a system of finite linear length $L_z$ limiting the length of the skyrmion string; here a threshold current $v^\parallel_{s,{\rm cr}} \sim \frac{\mathcal{D} \pi}{L_z |\beta/\alpha - 1|}$ will be required although surface twist and anchoring effects could lead to further complications \cite{Koshibae20}. 

Disorder, in particular, breaks the translational symmetry and induces finite threshold currents both for longitudinal currents, $v^\parallel_{s,{\rm cr}}$, as well as for the in-plane motion of skyrmion strings due to in-plane currents, $v^\perp_{s,{\rm cr}}$. In the following, we estimate this effect assuming collective pinning by weak disorder \cite{Blatter1994}. 
It can be shown \cite{SI} that the two threshold velocities for a skyrmion string lattice are related by $v_{s,\text{cr}}^\parallel\approx \sqrt{v_{s,\text{cr}}^\perp \eta \mathcal{D}/\xi}$, where $\eta = c_{44}/c_{66}$ is the ratio of the bend and shear elastic constants of the lattice, and $\xi$ is a length scale characterizing the disorder potential. As the size of the skyrmion is the smallest length scale that can be resolved by the string's elasticity we approximate $\xi \sim 2\pi/Q$.
From the dispersion relation for the spin waves, see Fig.~\ref{fig4}(a) and Ref.~\cite{Garst2017}, we can estimate $\mathcal{D} Q^2 \approx 0.5 \omega_{c2}$ as well as $\eta \approx 0.5$. 
Schulz {\it et al.}~\cite{Schulz2012} provide for a high-purity sample of MnSi a value for the critical in-plane velocity $v^\perp_{s,{\rm cr}} \sim 10^{-4}$ m/s of the skyrmion lattice motion induced by a critical charge current $j^\perp_{{\rm cr}} \sim 10^6$ A/m$^2$. Using $\omega_{c2}/(2\pi) = 16.7$ GHz and $2\pi/Q = 18$ nm for MnSi \cite{Garst2017}, we obtain the estimate for the longitudinal threshold velocity $v_{s,\text{cr}}^\parallel
\sim 0.03\, {\rm m/s}$, that is two orders of magnitudes larger than the transversal threshold. This amounts to a critical current $j^\parallel_{\rm cr} \approx \frac{v^\parallel_{s,{\rm cr}}}{v^\perp_{s,{\rm cr}}} j^\perp_{\rm cr} \sim 3\times 10^8 {\rm A/m}^2 $ for the Goldstone instability in MnSi highlighting the necessity of ultrapure samples.

In summary, a spin current flowing longitudinal to magnetic skyrmion strings leads to a Goldstone spin wave instability. Its application allows, in principle, to melt the skyrmion string lattice and induce a dynamical phase transition. Depending on the parameters, the resulting state could be, for example, a static polarized phase, a dynamical conical state with moving phase fronts \cite{Nagaosa2019,Yokouchi2020}, or other dynamically ordered and disordered phases \cite{Shibata2005,He2008,Lin2013}.

M.G. is supported by the Deutsche Forschungsgemeinschaft (DFG, German Research Foundation) via project-id 403030645 and project-id 445312953, and V. K. is partially supported by the Program of Fundamental Research of the Department of Physics and Astronomy of the National Academy of Sciences of Ukraine (Project No. 0120U100855).

\newpage

\begin{widetext}

{\bf \large \centering Supplementary Information\\}

\vspace{1em}

%
%
%
%


We provide details of our analytical and numerical analysis. In section \ref{app:Spectrum}, we present the derivation of the fluctuation spectrum that gives access to the stability criterion. In section \ref{app:magnon}, the magnon spectrum and instability diagram for a single skyrmion string at a dimensionless field $h = 1.08$ is provided. Section \ref{app:Pinning} discusses the pinning of skyrmion strings and gives estimates for the resulting threshold currents. Finally, section \ref{app:simuls} gives details of the micromagnetic simulations.

\end{widetext}



\section{Fluctuation spectrum of skyrmion strings in the presence of longitudinal spin currents and damping}
\label{app:Spectrum}

In this chapter we derive the  fluctuation spectrum of Eq.~(3) in the main text. The starting point is the Landau-Lifshitz-Gilbert equation given in Eq. (1) of the main text in the presence of a longitudinal spin current, i.e., $\vec{v}_s = v_s \hat{\vec z}$. It is convenient to consider this equation in the moving frame of reference, $\vec{r}'=\vec{r}-t\vec{v}_s\beta/\alpha$, $t'=t$, where it reduces to 
\begin{equation}\label{eq:LLG-mov}
	\left(\partial_{t'}-\tilde{v}_s\partial_{z'}\right)\vec{n}=-\gamma\vec{n}\times\vec{B}_{\text{eff}}+\alpha\vec{n}\times\partial_{t'}\vec{n},
\end{equation}
with $\tilde{v}_s=v_s(\beta/\alpha-1)$ and the effective field $\vec{B}_{\text{eff}} = - \frac{1}{M_s} \frac{\delta V}{\delta \vec{n}}$. We consider an equilibrium magnetic texture that only varies within the plane perpendicular to the $z$-axis, i.e., $\vec{n}_0 = \vec{n}_0(x',y')$, that is the case, for example, for the single skyrmion string or the skyrmion string lattice. Such a texture is static in the moving frame of reference, and it also solves the equation of motion in the presence of a finite $\tilde v_s$ as $\partial_{z'} \vec n_0 = 0$. 

In the following, we perform a linear stability analysis by considering small amplitude fluctuations around $\vec n_0$. Following Ref.~\cite{Schuette14}, we use the parametrization 
\begin{equation}\label{eq:}
	\vec{n}=\vec{n}_0\sqrt{1-2|\psi|^2}+\vec{e}^+\psi+\vec{e}^-\psi^*
\end{equation} 
in terms of the complex-valued wavefunction $\psi$ where $\vec{e}^\pm=(\vec{e}_1\pm i\vec{e}_2)/\sqrt{2}$ with $\vec{e}_1$ and $\vec{e}_2$ forming an orthonormal basis within a plane perpendicular to $\vec{n}_0$, such that $\vec{e}_1\times\vec{e}_2=\vec{n}_0$. Linearizing the equation of motion in $\psi$ we obtain the wave equation for the spinor $\vec \Psi = (\psi, \psi^*)^T$,
\begin{align}
i \tau^z (\partial_{t'} - \tilde v_s \partial_{z'}) \vec \Psi = \mathcal{H} \vec \Psi + \alpha \partial_{t'}  \vec \Psi,
\end{align}
with the Pauli matrix $\tau^z$ and the spin wave Hamiltonian $\mathcal{H}$. For the single skyrmion string, the explicit form of $\mathcal{H}$ was discussed in detail in Refs.~\cite{Lin2019,Kravchuk2020}. 

\begin{figure}[b]
	\includegraphics[width=\columnwidth]{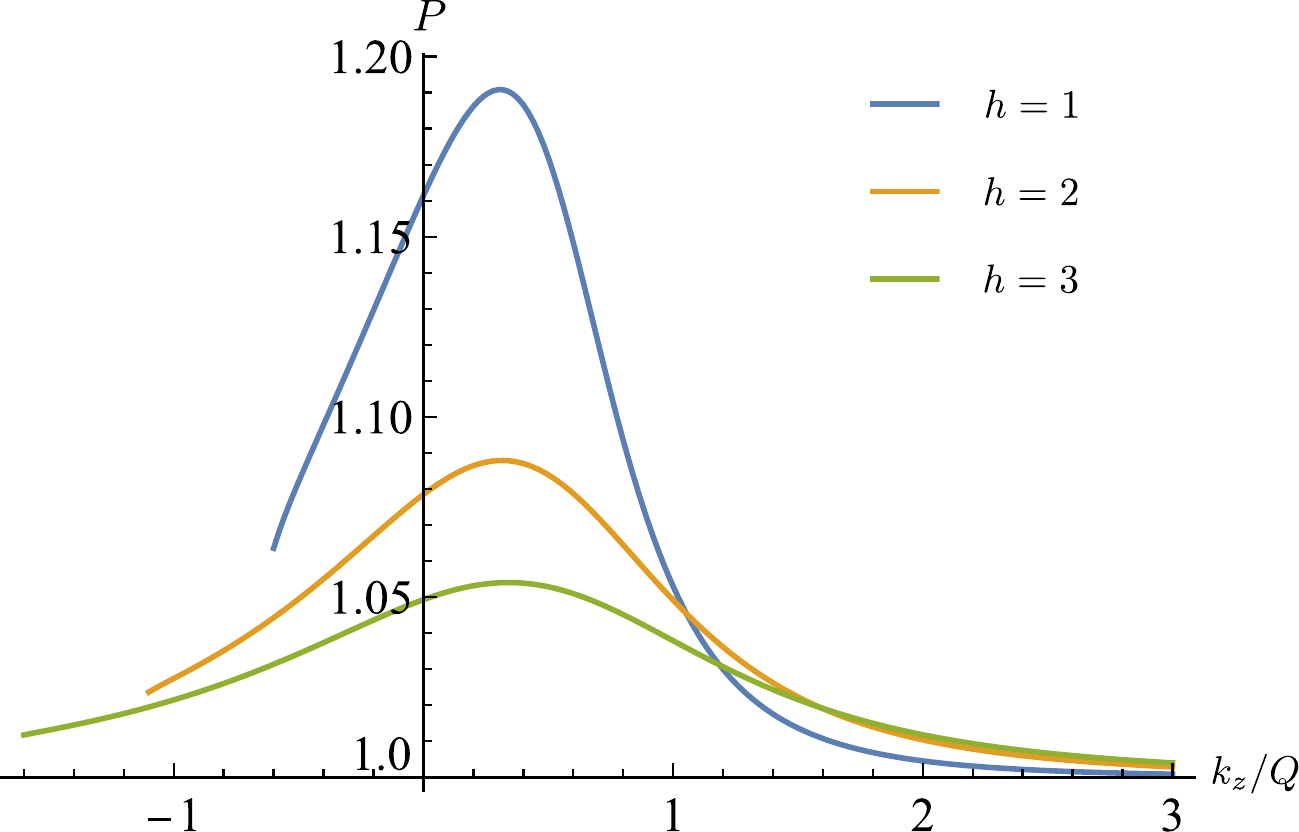}
	\caption{The ellipticity parameter $P(k_z)$ for the translational Goldstone mode of a single skyrmion string as a function of the wavevector along the string $k_z$ for various dimensionless magnetic fields $h$.}\label{fig:P}
\end{figure}

Performing a partial Fourier transformation $\vec \Psi(x', y', z',t') = \int \frac{d\omega}{2\pi} \frac{d k_z}{2\pi} e^{-i \omega t' + i k_z z'} \vec \Psi(x', y', k_z, \omega)$, the spin wave equation simplifies to 
\begin{align} \label{SWEq1}
 (\omega + \tilde v_s k_z) \tau^z \vec \Psi = \mathcal{H} \vec \Psi - i \alpha \omega  \vec \Psi,
\end{align}
where derivatives with respect to the $z'$ coordinate  in the Hamiltonian should be replaced $-i \partial_{z'} \to k_z$. The eigenfunctions of the Hamiltonian $\mathcal{H}$ fulfil the eigenvalue equation
\begin{align}
\mathcal{H} \vec \Psi_{n,k_z}(x',y') = \tau^z \omega_n(k_z) \vec \Psi_{n,k_z}(x',y'),
\end{align}
where the additional quantum number $n$ is either discrete or continuous, and the eigenfunctions obey the orthogonality relation 
\begin{align}
\int dx' dy' \vec \Psi^\dagger_{n,k_z}(x',y') \tau^z \vec \Psi_{n',k_z}(x',y') =
\delta_{n,n'}.
\end{align}
Here, $\delta_{n,n'}$ is either a Kronecker symbol for discrete quantum numbers $n$ and $n'$ or a delta function for continuous quantum numbers. Expanding Eq.~\eqref{SWEq1} in eigenfunction of the Hamiltonian and treating the damping $\alpha$ in lowest order perturbation theory, we obtain an equation for the frequency $\omega$,
\begin{align}
\omega + \tilde{v}_s k_z = \omega_n(k_z) - i \alpha \omega P_n(k_z),
\end{align}
where 
\begin{align} \label{Ellipt}
P_n(k_z) = \int dx' dy' \vec \Psi^\dagger_{n,k_z}(x',y') \vec \Psi_{n,k_z}(x',y').
\end{align}
Dropping the index $n$ and solving for $\omega$ we get in first order in $\alpha$
\begin{align}
\omega = (\omega(k_z) - \tilde{v}_s k_z) (1 - i \alpha P(k_z)).
\end{align}
The fluctuation spectrum $\omega_{v_s}(k_z)$ in the laboratory frame of reference given in Eq.~(3) of the main text is finally obtained after a shift, $\omega_{v_s}(k_z) = \omega + \frac{\beta}{\alpha} v_s k_z$, i.e., 
\begin{align} \label{spectrum}
&\omega_{v_s}(k_z) = \\\nonumber
&\frac{\beta}{\alpha} v_s k_z + \left(\omega(k_z) - v_s \left(\frac{\beta}{\alpha} -1\right) k_z\right) (1 - i \alpha P(k_z)).
\end{align}
Note that the spin-wave Doppler shift, $\frac{\partial}{\partial v_s}$  Re $\omega_{v_s}(k_z) = k_z$, is universal and independent of $\alpha$ and $\beta$ in agreement with previous reports \cite{Fernandez-rossier2004}.

The parameter $P(k_z)$ of Eq.~\eqref{Ellipt} measures the ellipticity of the mode \cite{Kambersky1975} that was discussed recently for skyrmions by R\'ozsa {\it et al.} \cite{Rozsa2018}. For a single skyrmion string, the dependence of $P(k_z)$ for the translational Goldstone mode is shown in Fig.~\ref{fig:P}.

\section{Spectrum and instability diagram for a single skyrmion string}
\label{app:magnon}

\begin{figure}
	\includegraphics[width=0.8\columnwidth]{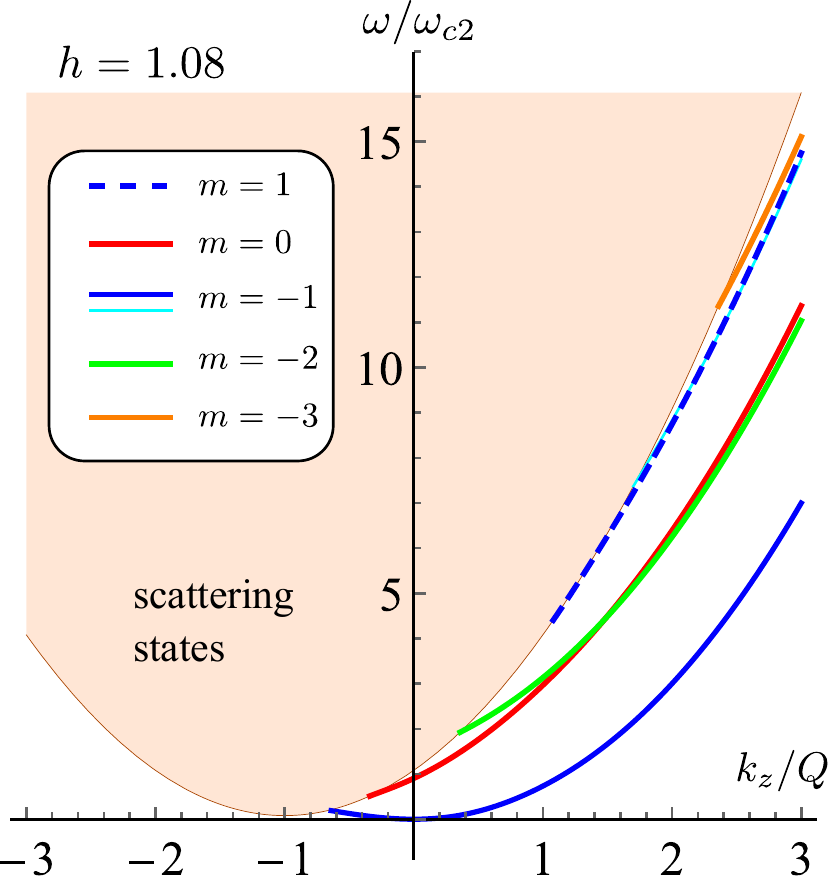}
	\caption{Linear spin wave spectrum of a single skyrmion string for the dimensionless field $h = 1.08$. Besides the scattering states, there exist various bound magnon-skyrmion states labeled by the quantum number $m$: translational Goldstone mode (blue line, $m=-1$), breathing mode (red line, $m = 0$), quadrupolar mode (green line, $m = -2$), higher-order $m=-1$ mode (light blue line), counterclockwise mode (dashed line, $m=1$), and sextupolar mode (orange line, $m=-3$).}
\label{fig:h108}
\end{figure}

\begin{figure}
\includegraphics[width=0.8\columnwidth]{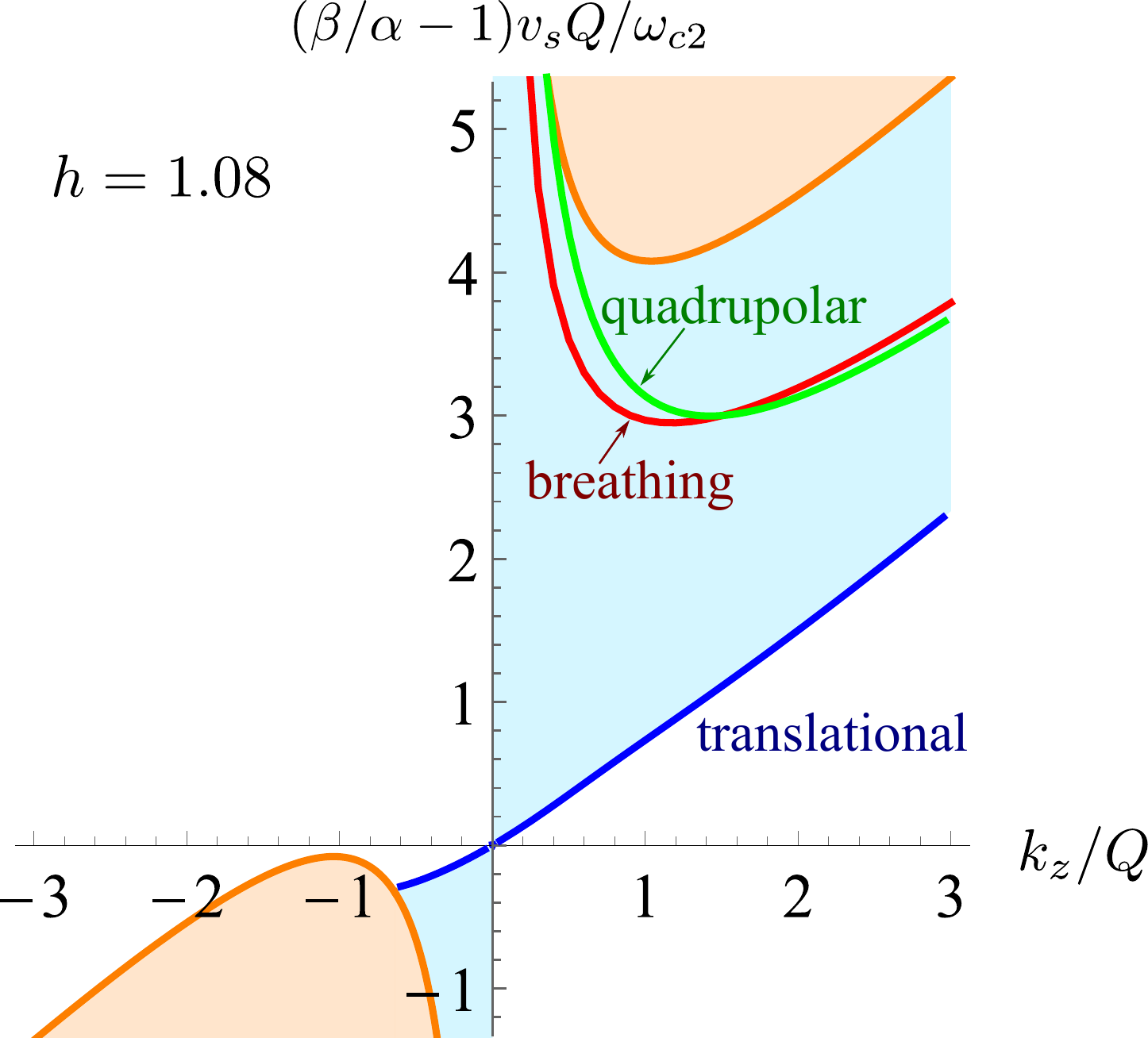}
\caption{The instability diagram for the single string at a dimensionless field $h = 1.08$ deriving from the spectrum in  Fig.~\ref{fig:h108} and the stability criterion of Eq.~(4) in the main text. The translational Goldstone mode is unstable in the light-blue region. In addition, the breathing mode and the quadrupolar mode are becoming unstable at the red and green lines, respectively. The field-polarized ground state is unstable within the orange shaded area.
}\label{fig:stab-B04}
\end{figure}

Using the spectrum $\omega(k_z)$ for a single skyrmion string, the instability diagram can be constructed with the help of Eq.~\eqref{spectrum} as discussed in the main text for a dimensionless magnetic field $h = 2.16$. Here, we present the spectrum and the resulting instability diagram in Fig.~\ref{fig:h108} and \ref{fig:stab-B04}, respectively, for $h = 1.08$ just above the transition to the conical state at $h_{c2} = 1$. For lower fields, additional modes appear below the spin-wave gap as specified in the figure captions.

\section{Collective pinning theory for threshold currents}
\label{app:Pinning}

Projecting the Landau-Lifshitz-Gilbert equation in the presence of a spin current, Eq.~(1) of the main text, onto the translational degree of freedom of the skyrmion texture, the Thiele equation \cite{Thiele1973} for its collective coordinate $\vec{u}$ is obtained,
\begin{align}
\label{eq:Thiele}
&\vec{G}\times\Big(\partial_t{\vec{u}}-\vec{v}^\perp_s+v^{\parallel}_s\partial_z\vec{u}\Big) + 
\nonumber\\&\qquad
\vec{D}_{\rm diss} \Big(\alpha\partial_t{\vec{u}}-\beta\vec{v}^\perp_s+\beta v^{\parallel}_s\partial_z\vec{u}\Big)
=\vec{f}.
\end{align}
The gyrocoupling vector is $\vec{G}= - \hat{\vec{z}} 4\pi M_s/\gamma$ and $\vec{D}_{\rm diss}$ is the dissipative tensor that can be approximated to be diagonal $D^{\rm diss}_{ij} = D_{\rm diss} (\delta_{ij} - \hat z_i \hat{z}_j)$ with a finite component only within the $(x,y)$ plane of the texture \cite{Everschor2011}. We decomposed the spin current into a longitudinal and a transversal component, $\vec{v}_s = \vec{v}^\perp_s + v^{\parallel}_s \hat{\vec{z}}$ with $\vec{v}^{\perp}_s \hat{\vec z} = 0$. The collective coordinate $\vec{u}$ can be interpreted as a displacement field. For a single skyrmion string, it depends on time and the $z$-component, $\vec{u} = \vec{u}(z,t)$. For a skyrmion lattice, the displacement field in addition depends on the $x$ and $y$ coordinates, $\vec{u} = \vec{u}(x,y,z,t)$, and describes the small amplitude deviations of skyrmions from their equilibrium positions within each magnetic unit cell. The force per length of the skyrmion strings $\vec{f}$ on the right hand side of Eq.~\eqref{eq:Thiele} is attributed both to the elasticity of the skyrmion texture as well as to the influence of impurities. 

In the absence of disorder, the force arising from the displacement field $\vec{u}$
reduces for small deviations to $\vec{f} = - \mathcal{D} |G| \partial_z^2 \vec{u}$ such that for $\vec{v}_s = 0$ and $\alpha = 0$ the low-energy limit of the Goldstone spectrum is recovered, $\omega(k_z) \approx \mathcal{D} k^2$, where $\mathcal{D}$ is the stiffness of the Goldstone mode. The product $\mathcal{D} |G|$ thus corresponds to the elastic constant of skyrmion strings. Note that the numerical value of  $\mathcal{D}$ differs for the single skyrmion string and the string lattice. 

The aim of this section is to estimate the threshold current for the skyrmion motion in the presence of disorder in case of an in-plane current $\vec{v}^\perp_s$ as well as for the instability induced by the longitudinal current $v^{\parallel}_s$. We consider the competition between the elasticity of the texture and the impurities within the framework of  collective pinning theory implicitly assuming that the impurity potential is sufficiently weak. In this limit, the properties of $\vec{f}$ share similarities with the case of Abrikosov vortices in type II superconductors so that we can apply the results developed for the latter case \cite{Blatter1994}. In the following, using arguments of Ref.~\cite{Blatter1994} we derive estimates of the order of magnitude of the threshold currents for the skyrmion string as well as for the skyrmion string lattice.

\subsection{Collective pinning of a single skyrmion string}

An important length scale characterising the competition between the string's elasticity and the disorder is the collective pinning length $L_{\text{pin}}$ that quantifies the length of a string segment at which its displacement becomes on the same order as the typical length $\xi$ characterizing the disorder potential. 
%
%
For a dimensional estimate we integrate the Thiele equation over a skyrmion string segment of length $L_{\text{pin}}$ and  obtain in the pinned regime with $\partial_t{\vec{u}} = 0$,
%
\begin{align}
\label{eq:Thiele2}
L_{\text{pin}} \vec{G}\times\Big(-\vec{v}^\perp_s+v^{\parallel}_s 
\partial_z\vec{u}
\Big) \approx \vec{F}_{\text{pin}}.
\end{align}
Here, we neglected terms proportional to small factors of $\alpha$ and $\beta$ for an order of magnitude estimate. Using the estimate $|\partial_z \vec{u}| \sim \xi/L_{\text{pin}}$ as well as the result for the pinning force given in Ref.~\cite{Blatter1994}, $F_{\text{pin}} = |\vec{F}_{\text{pin}}| \sim \frac{\mathcal{D} |G| \xi}{L_{\text{pin}}}$, we obtain the following estimates for the threshold currents,
\begin{equation}\label{eq:vc}
	v_{s,\text{cr}}^\perp \sim \frac{\mathcal{D}\xi}{L_{\text{pin}}^2},\qquad 
	v_{s,\text{cr}}^\parallel \sim \frac{\mathcal{D}}{L_{\text{pin}}}.
\end{equation}
As the longitudinal current couples to the derivative of the displacement field, its threshold current is suppressed compared to the threshold current for the in-plane motion, 
$v_{s,\text{cr}}^\perp/v_{s,\text{cr}}^\parallel \sim \xi/L_{\text{pin}} \ll 1$, because the pinning length is much larger than $\xi$ for weak pinning. Eliminating the pinning length we can relate the two threshold currents, 

\begin{align}\label{eq:vpar-single}
	v_{s,\text{cr}}^\parallel \sim \sqrt{v_{s,\text{cr}}^\perp\mathcal{D}/\xi}.
\end{align}

\subsection{Collective pinning of a skyrmion string lattice}

In general, the collective pinning theory of Abrikosov vortex lattices is quite involved due to the presence of several length scales leading to various regimes. In Ref.~\cite{Blatter1994}, assuming point-like impurities the disorder length $\xi$ is basically identified with the size of the vortex core. Taking over this assumption for the skyrmion strings, $\xi$ is on the order of the skyrmion size $r_s$. In the field range where the skyrmion lattice is thermodynamically stable, its lattice constant $a_0$ is on the same order as the skyrmion size $r_s$ and thus $\xi \sim a_0$. We will assume this relation in the following which simplifies the discussion considerably.

For the skyrmion string lattice, a collective pinning volume $V_{\text{pin}}=\ell_{\text{pin}}\varrho_{\text{pin}}^2$ is considered, where $\ell_{\text{pin}}$ and $\varrho_{\text{pin}}$ are longitudinal, i.e., along the string, and transversal dimensions of the pinning volume domain, respectively. They are given by \cite{Blatter1994}
\begin{equation}\label{eq:L-R}
	\ell_{\text{pin}}\approx \eta\frac{L_{\text{pin}}^3}{a_0^2},\qquad \varrho_{\text{pin}}\approx\sqrt{\eta}\frac{L_{\text{pin}}^3}{a_0^2},
\end{equation}
where $\eta=c_{44}/c_{66}$ is the ratio of the bend ($c_{44}$) and shear ($c_{66}$)  elastic constants of the skyrmion lattice. 
The collective dynamics of the pinned volume element $V_{\text{pin}}$ is governed by the equation
\begin{equation}\label{eq:Thiele-Vc}
	\vec{G}^V\times\left(-\vec{v}^\perp_s+v^{\parallel}_s\partial_z\vec{u}\right)=\vec{F}_{\text{pin}}^V,
\end{equation}
where $\vec{G}^V=\vec{G}\ell_{\text{pin}}\varrho_{\text{pin}}^2/a_0^2$ is the net gyrovector of $V_{\text{pin}}$. The pinning force acting on the collective pinning volume $V_{\text{pin}}$ is related to the pinning force $F_{\text{pin}}$ acting on the segment $L_{\text{pin}}$ of the single string \cite{Blatter1994},
\begin{equation}\label{eq:FpinV}
	F_{\text{pin}}^V \sim F_{\text{pin}}\eta\frac{L_{\text{pin}}^4}{a_0^4}.
\end{equation}
With the estimate $|\partial_z\vec{u}|\sim\xi/\ell_{\text{pin}}$ we obtain for the threshold velocities
\begin{equation}\label{eq:vc-lat}
v_{s,\text{cr}}^\perp\approx\frac{\mathcal{D}\xi}{\eta}\frac{a_0^4}{L_{\text{pin}}^6},\qquad 
v_{s,\text{cr}}^\parallel\approx\mathcal{D}\frac{a_0^2}{L_{\text{pin}}^3},
\end{equation}
that obey the relation
\begin{equation}\label{eq:vpar-lat}
v_{s,\text{cr}}^\parallel\approx \sqrt{v_{s,\text{cr}}^\perp \eta \mathcal{D}/\xi}.
\end{equation}
Formally, the result is similar to the one of the single string \eqref{eq:vpar-single} except for the additional factor $\eta=c_{44}/c_{66}$ characterizing the elasticity of the string lattice and that the numerical value of $\mathcal{D}$ differs in the two cases.

\subsection{Discussion of threshold velocities for the skyrmion string lattice in MnSi}

In the following, we further discuss the result \eqref{eq:vpar-lat} for the skyrmion string lattice and give numerical estimates. The lattice constant for the triangular lattice $a_0 = \frac{4\pi}{\sqrt{3} k_{\rm SkL}}$ can be related to the primitive reciprocal lattice vector that is approximately $k_{\rm SkL} \approx Q$ in the field range where the skyrmion lattice is thermodynamically stable. From the dispersion relation for the spin waves, see Fig.~4(a) of the main text and Ref.~\cite{Garst2017}, we can estimate $\mathcal{D} Q^2 \approx 0.5 \omega_{c2}$ as well as $\eta \approx 0.5$. 

More specifically, for MnSi the typical frequency and wavevectors are $\omega_{c2}/(2\pi) = 16.7$ GHz and $2\pi/Q = 18$ nm \cite{Garst2017} so that the effective velocity $\eta\mathcal{D}/a_0 \sim 10$ m/s.
Schulz {\it et al.}~\cite{Schulz2012} provide for a high-purity sample of MnSi a value for the critical in-plane velocity $v^\perp_{s,{\rm cr}} \sim 10^{-4}$ m/s of the skyrmion lattice motion induced by a critical charge current $j^\perp_{{\rm cr}} \sim 10^6$ A/m$^2$. This leads to the estimate for the longitudinal threshold velocity using $\xi \sim a_0$,
\begin{align}
v_{s,\text{cr}}^\parallel
\sim 0.03\, {\rm m/s}.
\end{align}
This amounts to a critical current $j^\parallel_{\rm cr} \approx \frac{v^\parallel_{s,{\rm cr}}}{v^\perp_{s,{\rm cr}}} j^\perp_{\rm cr} \sim  3\times 10^8 {\rm A/m}^2 $ for the Goldstone instability in MnSi.

\section{Details of micromagnetic simulations}
\label{app:simuls}

In this section, we provide details of the performed micromagnetic simulations used to generate the results of Figs.~1, 3, and 4 in the main text. For the simulations, we used the software Mumax3~\cite{Vansteenkiste14} to determine the time evolution according to the Landau-Lifshitz-Gilbert equation in presence of spin-currents, see Eq.~(1) from the main text.

For concreteness, we chose the material parameters $A=8.78$~pJ/m, saturation magnetization $M_s= 0.384$~MA/m ($\mu_0 M_s=0.48$ T), DMI constant $D=1.58$~mJ/m$^2$ \cite{Beg15}, and the values for the Gilbert damping and nonadiabaticity parameters $\alpha=0.0036$ and $\beta=0.01$, respectively. The chosen parameters imply the following values for the saturation field  $\mu_0H_{c2}=D^2/(2AM_s)= 0.37$~T, the frequency  $\omega_{c2}=\gamma\mu_0H_{c2}=2\pi/(96\,\text{ps}) = 2\pi \, 10.4$ GHz, the wave-vector $Q=D/(2A)=2\pi/(70\,\text{nm})$, and the velocity  $v_0=\omega_{c2}/Q=0.724$~km/s. The simulations were carried out in the presence as well as in the absence of long-range magnetostatic interactions without changing the main results.

In order to determine the position of the skyrmion string $\vec{R}(z,t)=X_1(z,t)\hat{\vec{x}}+X_2(z,t)\hat{\vec{y}}$ from the simulation data, we employed the definition in terms of the first moment of topological charge~\cite{Kravchuk2020}
\begin{equation}\label{eq}
	X_i(z,t)=\frac{1}{N_{\text{top}}}\int x_i\rho_{\text{top}}(x,y,z)\md x\md y,
\end{equation}
where $(x_1,x_2)=(x,y)$,  $\rho_{\text{top}}(x,y)=\frac{1}{4\pi}\vec{n}\cdot[\partial_{x}\vec{n}\times\partial_y\vec{n}]$ is topological charge density and $N_{\text{top}}=\int\rho_{\text{top}}\md x\md y$ is the total topological charge. Note that $N_{\text{top}}$ is independent of $z$ and $t$. 

For the simulations with a single skyrmion string (Figs.~1 and 3 in the main text), we considered a rectangular simulation sample with size $L_x=L_y=100$~nm and $L_z=700$~nm. The number of moments is $N=N_x\times N_y\times N_z$, where $N_x=N_y=100$ and $N_z=256$.
The simulations were performed with fixed boundary conditions in $x$- and $y$-direction and periodic boundary conditions in the $z$-direction.

In order to prepare the initial state for Fig.~1 of the main text we relaxed the skyrmion string in a magnetic field $\vec{H}=H\hat{\vec{z}}+\vec{h}_{\text{rnd}}$ with a small random component $\vec{h}_{\text{rnd}}$ with constant amplitude $\mu_0|\vec{h}_{\text{rnd}}|=5$~mT but random direction for each discretization unit cell. The averaged string position coincides with the sample center, $\langle\vec{R}(z)\rangle_z\equiv(0,0)$; the averaged deviation is $\langle|\vec{R}(z)|\rangle_z=0.01$~nm.

In order to obtain the helix-shaped string as an initial state for Fig.~3 of the main text, we first relaxed the skyrmion string in a uniform field $\vec{H}=H\hat{\vec{z}}$. Afterwards, we applied a space-dependent magnetic field of the form $\tilde{\vec{H}}=[H+\delta H_z(x,y,z)]\hat{\vec{z}}$ with $\delta H_z(x,y,z)=\tilde{H}_z[x\cos(zk_z)/L_x+y\sin(zk_z)/L_y]$, and, again, let the string relax. The amplitude $\tilde{H}_z$ is set at a particular value depending on $k_z$ so that the initial helix-wave radius becomes approximately $R_0\sim1$~nm. The constant $H$ was chosen $\mu_0H=0.8$~T for Figs.~1 and 3 of the main text, and $\mu_0H=0.2$~T for Fig.~4 of the main text.

In Fig.~3(a,c) we use the Fourier transform amplitude $\mathcal{A}_k(t)=|\hat{\Psi}_k(t)|$, where
\begin{equation}\label{eq:RF}
	\hat{\Psi}_k(t)=\frac{1}{L_z}\int\limits_0^{L_z}\Psi e^{-ikz}\md z, \qquad \Psi=X_1+iX_2.
\end{equation}
Note that for a helix-shaped string $\Psi=R_0e^{ik_0z}$ we have $\mathcal{A}_k=R_0\delta_{k,k_0}$.

Figs.~3(b,c) show the dependence of the helix-wave radius $R_0=\mathcal{A}_{k_z}$ on time for a fixed wave-vector $k_z/Q=1$. The growth of $R_0$ becomes slower as $v_s$ decreases and $R_0$ decays for currents smaller than a critical value $\tilde{v}_s^*$. Fitting the numerically evaluated $R_0(t)$ by an exponential function $a+b e^{c t}$ for several $\tilde{v}_s$, we determine the dependence of the exponent on the velocity, $c = c(\tilde{v}_s)$ by means of linear interpolation. The critical current $\tilde{v}_s^*$ is obtained by solving $c(\tilde{v}^*_s)=0$. The resulting values of $\tilde{v}_s^*$ for several $k_z$ are shown as yellow dots in Fig.~3(a) of the main text. 

For the simulation of the lattice of skyrmion strings shown in Fig.~4(d) of the main text, we considered a rectangular non-primitive unit cell of the triangular lattice containing two skyrmion strings. Periodic boundary conditions were applied along all three directions. The cell size $L_x=80$~nm, $L_y=139$~nm was chosen such that the total energy per unit cell is minimized for the field $\mu_0 H=0.2$~T applied in $z$-direction. The sample size along the $z$-direction was $L_z=700$~nm; the discretization parameters were $N_x=100$, $N_y=174$, and $N_z=256$.
The position of the two skyrmions within the non-primitive unit cell were individually determined with the help of Eq.~\eqref{eq} by properly limiting the integration range. In our simulations, both skyrmion positions were found to oscillate in phase. Similar to the case of a single string, we generated an initial state with a helical perturbation using a 
space-dependent magnetic field, and we estimated $\tilde{v}_s^*$ in Fig.~4(d) of the main text from simulations for various values of $\tilde{v}_s$ and $k_z$.


\bibliography{SkStringInstability}

\end{document}